\documentstyle[preprint,aps]{revtex}
\begin{document}
\draft
\preprint{}
\title{ Neutrino Flavor Mixing \\ 
in the `Doublet-Singlet Oscillation' Model}
\author{ Nobuchika Okada 
 \thanks{e-mail: n-okada@phys.metro-u.ac.jp}
\thanks{JSPS Research Fellow}}
\address{Department of Physics, Tokyo Metropolitan University,\\
         Hachioji-shi, Tokyo 192-03, Japan}
%\author{and}
\date{\today}
\preprint{
\parbox{4cm}{
\baselineskip=12pt
TMUP-HEL-9607\\
June, 1996\\
\hspace*{1cm}
}}
\maketitle
\vskip 2.5cm
\begin{center}
{\large Abstract}
\vskip 1.0cm
\begin{minipage}[t]{14cm}
\baselineskip=19pt
\hskip4mm
%%%%%%%%%%%%%%%%%
%   abstract    %
%%%%%%%%%%%%%%%%%
Recently, it was pointed out that the solar and atmospheric neutrino 
deficits can be explained by oscillations between electroweak doublet and singlet 
neutrinos in the model of Majorana neutrinos.   
However, since the model includes no flavor mixing, it cannot explain the recent 
LSND result.  
We extend the model to include the flavor mixing, 
and obtain the explanation of the LSND result 
together with  the solar and atmospheric neutrino deficits.  
The requirement for the neutrinos to be the hot dark matter 
selects out only the vacuum oscillation solution to the solar neutrino deficit. 
%%%%%%%%%%%%%%%
\end{minipage}
\end{center}
\newpage
%%%%%%%%%%%%%%%%%%%%%%%%%%%%%%
\def\barr{\begin{eqnarray}}
\def\earr{\end{eqnarray}}
%%%%%%%%%%%%%%%%%%%%%%%%%%%%%%
The recent LSND experiment \cite{lsnd1} may be the first observation
of the neutrino oscillation 
in $\overline{\nu_{\mu}} \rightarrow \overline{\nu_{e}}$ conversion.
The experiment obtained values of parameters of the oscillation 
as $\Delta m_{{}_{LSND}}^2\simeq (0.25$--$2.5$) ${\rm{eV}}^2$ and 
$\sin^22\theta_{{}_{LSND}} \simeq 2\times 10^{-3}$ -- $4 \times 10^{-2}$.
However, the LSND data is still somewhat controversial:
in ref.[2] it is interpreted as an upper bound on the parameters
of the neutrino oscillation. 
In this latter, we accept the data as the evidence of the neutrino oscillation.
Then, we should consider the LSND result
together with another neutrino oscillation
phenomena, i.e. the solar neutrino deficit \cite{solar}
and the atmospheric neutrino anomaly \cite{atm}.

Recently, it was pointed out that 
the solar and atmospheric neutrino deficits can be explained by
oscillations between electroweak doublet and singlet neutrinos
in the model of the Majorana neutrino without flavor mixing 
\cite{dsoscillation}.
In the following, this type of neutrino oscillation
is called `doublet-singlet oscillation'.
The solar electron neutrino deficits can be explained
by the `doublet-singlet oscillation' in the first generation 
(two solutions are possible for the values of 
$\Delta m^2_{\odot}$ and $\sin^22\theta_{\odot}$ 
as they will be mentioned in the following), 
and the atmospheric muon neutrino anomaly, which requires 
$\Delta m^2_{\oplus}\simeq 10^{-2}$ ${\rm{eV}}^2$ and 
$\sin^22\theta_{\oplus}\simeq 1$, can be done by 
the `doublet-singlet oscillation' in the second generation. 
Furthermore, neutrinos can be both two components of the dark matter
in the cold plus hot dark matter cosmological models \cite{chdm}.
While some neutrinos in the first and second generation
can be the hot dark matter, tau neutrino can be the cold dark matter.
There are two kinds of solutions to the hot dark matter neutrino.
One is that only two neutrinos 
in the second generation are the hot dark matter,
if we choose a solution called the `small-angle MSW solution' \cite{msw} 
($\Delta m^2_{\odot}\simeq 10^{-5}$ ${\rm{eV}}^2$ and
$\sin^22 \theta_{\odot}\simeq10^{-2}$) in the first generation.
The other is that all of the neutrinos in two generations are
the hot dark matter with the choice 
of the vacuum oscillation solution \cite{vacuum}
($\Delta m^2_{\odot}\simeq 10^{-10}$ ${\rm{eV}}^2$ and
$\sin^22 \theta_{\odot}\simeq1$).  

However, it is clear that this model cannot explain  
the $\overline{\nu_{e}}$ appearance from $\overline{\nu_{\mu}}$ 
in the LSND experiment, 
since there is no flavor mixing.
In this latter, we extend the model in ref.\cite{dsoscillation}
to include flavor mixing between the first and second generations 
\footnote{
Our model is the same, in form, as the models in ref.\cite{sterile}, 
which treat four neutrinos; three `active' neutrinos, 
$\nu_e$, $\nu_{\mu}$ and $\nu_{\tau}$, and one `sterile' neutrino.  
}. 
We will show that this extension selects only one
out of two solutions to the hot dark matter neutrino.

We extend the standard model by introducing three right-handed neutrinos
and one electroweak singlet scalar $\Phi$ \cite{cmp}.
The Yukawa couplings are described by
%%%%%
\barr
{\cal L}_{\rm{Yukawa}}=-\overline{\nu _{L }}(g_{{}_Y}\phi) \nu _{R }
-\overline{\nu_{R}{}^c}(g_{{}_M }\Phi) \nu_{R} +h.c. \; \; ,\label{yukawa}
\earr
%%%%%
where $\phi$ is the electric-charge neutral component of
the Higgs field in the standard model,
$g_{{}_Y}$ and $g_{{}_M }$ are $3\times3$ matrices, and $\nu _{L,R }$ are
column vectors.
The Dirac and Majorana mass terms appear by
the non-zero vacuum expectation values of above scalar fields.
To our aim, we focus the first and second generations.
Then the mass matrix is given by
%%%%%%%
\barr
\left[ \begin{array}{cc}
\hspace{0.4cm} 0 \hspace{0.3cm}  & \hspace{0.3cm} m_{{}_D} \hspace{0.4cm} \\
\hspace{0.4cm} m_{{}_D}^T \hspace{0.3cm}  & \hspace{0.3cm} M \hspace{0.4cm}
\end{array}\right]
\label{mat}\; ,
\earr
%%%%%%%%
where $m_{{}_D}=g_{{}_Y}\langle \phi \rangle$ is the $2\times2$ Dirac mass matrix,
and $M=g_{{}_M}\langle \Phi \rangle$
is the $2\times2$ Majorana mass matrix.
In order to discuss neutrino oscillations, we should diagonalize
the $4\times4$ mass matrix in eq.(2).
This diagonalization is complicated in general.
However, to our aim, it is satisfactory to consider a simple case.
We assume that two matrices $m_{{}_D}$ and $M$ can be simultaneously diagonalized:  
$U_L^{\dagger}m_{{}_{D}}U_R=diag[m_1^{{}_D},m_2^{{}_D}]$ and 
$U_R^T M U_R=diag[M_1,M_2]$, 
where $U_L$ and $U_R$ are the $2\times 2$ orthogonal matrix 
\footnote{
In this letter, we neglect CP violation phases.
} 
for rotations of left and right-handed states, respectively.

Without loss of generality, we start from $U_R=1$.
Denoting the left-handed flavor eigenstate and mass eigenstate
as $\Psi_{f}=(\nu_{eL}, \nu_{\mu L}, \nu_{eR}{}^c, \nu_{\mu R}{}^c)^T$ and
$\Psi_{m}=(\nu_1, \nu_2, \nu_3, \nu_4)^T$, respectively,
the mixing matrix is given by
%%%%%%%%
%%%%%%%%
\barr 
\Psi_m=\left[ \begin{array}{cc}
U_L &0 \\
0 & 1 \\\end{array}\right]
\left[ \begin{array}{cccc}
C_{\odot}  & 0 & S_{\odot}&0  \\
0 & C_{\oplus}  & 0 & S_{\oplus}  \\
-S_{\odot}  & 0 & C_{\odot}&0  \\
0 & -S_{\oplus}  & 0 & C_{\oplus}  \\
\end{array}\right]\Psi_f
\label{mix}\; ,
\earr
%%%%%%%%
where $C_{\odot, \oplus}$ and $S_{\odot, \oplus}$ denote
$\cos \theta_{\odot, \oplus}$ and $\sin \theta_{\odot, \oplus}$, respectively.
Note that the model in ref.\cite{dsoscillation} appear 
as the limit $U_L\rightarrow 1$.  
In this case, the solar and atmospheric neutrino deficits are explained
with $\nu_{eL} \rightarrow \nu_{eR}{}^c$ and
$\nu_{\mu L} \rightarrow \nu_{\mu R}{}^c$ conversions, respectively.
The $2\times 2$  orthogonal matrix $U_L$ is described by
%%%%%%%%
\barr
U_L= \left[ \begin{array}{cc}
C_{{}_{LSND}} & S_{{}_{LSND}} \\
-S_{{}_{LSND}}& C_{{}_{LSND}} \\
\end{array}\right] \; ,
\earr
%%%%%%%
where $C_{{}_{LSND}}$ and $S_{{}_{LSND}}$ denote
$\cos \theta_{{}_{LSND}}$ and $\sin \theta_{{}_{LSND}}$, respectively.
The matrix $U_L$ corresponds to the flavor mixing
between two electroweak doublet neutrinos.

Then, we calculate neutrino oscillation probabilities.
Note that the flavor mixing angle required by the LSND experiment is small 
$S_{{}_{LSND}}\ll 1$, and the three mass square differences required by
the solar and atmospheric neutrino deficits and the LSND experiment 
are hierarchical, namely, 
$\Delta m^2_{\odot} \ll \Delta m^2_{\oplus} \ll \Delta m^2_{{}_{LSND}}$. 
These features reduce our analysis of the neutrino oscillation 
from four flavor case to combinations of two flavor cases. 
We define three mass square differences as 
$\Delta m^2_{\odot}=m_3^2-m_1^2$, $\Delta m^2_{\oplus}=m_4^2-m_2^2$, 
and $\Delta m^2_{{}_{LSND}}=|m_1^2-m_2^2|$, 
where $m_i$ is the Majorana mass eigenvalue of $\nu_i$ ($i=1,2,3,4$).  

The probability of $\nu_{\mu} \rightarrow \nu_e$ conversion 
which corresponds to the LSND experiment is given by
%%%%%%%%%%%%%%
\barr
P_{\nu_{\mu} \rightarrow \nu_e} 
&\simeq &  \sin^2 2\theta_{{}_{LSND}} \sin^2
\left(\frac{\Delta m^2_{{}_{LSND}}L}{4E}\right)
-\frac{1}{4} \sin^22\theta_{{}_{LSND}} \sin^22\theta_{\odot}
\sin^2 \left(\frac{\Delta m^2_{\odot}L}{4E}\right)\\ \nonumber
&-& \frac{1}{4}\sin^22\theta_{{}_{LSND}} \sin^22\theta_{\oplus}
\sin^2\left(\frac{\Delta m^2_{\oplus}L}{4E}\right)
\\  \nonumber
&\simeq & \sin^2 2\theta_{{}_{LSND}} \sin^2 \left(\frac{\Delta m^2_{{}_{LSND}}L}{4E}
\right)
\; .
\earr
%%%%%%%%%%%%%%%
Here, we use 
$\sin(\Delta m^2_{\odot}L/4E) \ll 1 $ and $\sin(\Delta m^2_{\oplus}L/4E) \ll 1$ 
at the energy range ($E=(36$--$60$) MeV) and 
the oscillation length ($L=30$m) in the LSND experiment 
and the hierarchy among the three mass square differences. 

The survival probability of $\nu_e \rightarrow \nu_e $ which corresponds to
the solar neutrino deficit is given by
%%%%%%%%%%%%%%
\barr
P_{\nu_e \rightarrow \nu_e}
&\simeq &  C_{{}_{LSND}}^4 \left[1-\sin^2 2\theta_{\odot} \sin^2
\left(\frac{\Delta m^2_{\odot}L}{4E}\right)\right]
+S_{{}_{LSND}}^4\left[1- \sin^22\theta_{\oplus}
\sin^2 \left(\frac{\Delta m^2_{\oplus}L}{4E}\right)\right] \\ \nonumber
&+&\frac{1}{2}\sin^22\theta_{{}_{LSND}}\cos \left(\frac{\Delta m^2_{{}_{LSND}}L}{2E}
\right)  \; ,
\earr
%%%%%%%%%%%%%%%
where the hierarchy among the three mass square differences is used. 
Since $S_{{}_{LSND}} \ll 1$, 
the second term in the right-hand side in eq.(6) can be ignored. 
Note that there is no resonance $\nu_e\rightarrow \nu_{\mu}$ conversion 
by the matter effect in the sun \cite{msw}, 
because of the large value of $\Delta m^2_{{}_{LSND}}$.  
Thus, the third term can be also ignored, and we can reduce 
eq.(6) to 
%%%%%%%%%%%%
\barr
P_{\nu_e \rightarrow \nu_e}\simeq  
1-\sin^2 2\theta_{\odot} \sin^2 \left(\frac{\Delta m^2_{\odot}L}{4E}\right)
= 1-P_{\nu_e \rightarrow \nu_e^c} \; ,
\earr
%%%%%%%%%%%%
where we use $S_{{}_{LSND}} \ll 1$, 
and $P_{\nu_e \rightarrow \nu_e^c}$ is the conversion probability
from $\nu_{e L}$ into its singlet partner $\nu_{e R}{}^c$. 

Finally, the survival probability of $\nu_{\mu} \rightarrow \nu_{\mu}$ 
which corresponds to the atmospheric neutrino deficit is given by the exchange 
$\odot \leftrightarrow \oplus$ in eqs.(6) and (7), that is, 
%%%%%%%%%%%%%%
\barr
P_{\nu_{\mu} \rightarrow \nu_{\mu}}
\simeq  1-\sin^2 2\theta_{\oplus} \sin^2 
\left(\frac{\Delta m^2_{\oplus}L}{4E}\right)
= 1- P_{\nu_{\mu} \rightarrow \nu_{\mu}^c}\; .
\earr
%%%%%%%%%%%%%
Here $S_{{}_{LSND}} \ll 1$ and the hierarchy are also used,
and $P_{\nu_{\mu} \rightarrow \nu_{\mu}{}^c}$
is the conversion probability
from $\nu_{\mu L}$ into its singlet partner $\nu_{\mu R}{}^c$.
The flavor mixing due to $U_L$ little affect the analysis
of the solar and atmospheric neutrino deficits. 

Let us discuss the mass spectrum of the neutrinos.  
Because of the hierarchy among the three mass square differences, 
$\Delta m^2_{\odot} \ll \Delta m^2_{\oplus} \ll \Delta m^2_{{}_{LSND}}$, 
only two types of mass spectrum are possible: 
type (i) $m_1 < m_3< m_2 < m_4$ with 
$\Delta m^2_{{}_{LSND}}=m^2_2-m^2_1$, 
type (ii) $m_2 < m_4 < m_1 < m_3$ 
with $\Delta m^2_{{}_{LSND}}=m^2_1-m^2_2$.  

The value of $m_i$ is fixed by the solution to the hot dark matter 
neutrino.  
The sum of the neutrino masses is required $M_{HOT}=(5$--$7$) eV   
in the cold plus hot dark matter models \cite{chdm}. 
Then, we obtain 
%%%%%%%%%%
\barr
M_{HOT}=m_1+m_2+m_4 \simeq 
m_1 +2\; \sqrt{m_1^2 \pm \Delta m^2_{{}_{LSND}}}  \; , 
\earr
%%%%%%%%%%
where the hierarchy $\Delta m^2_{\oplus} \ll\Delta m^2_{{}_{LSND}}$ is used,  
and the signs $+$ and $-$ correspond to the type (i) and 
type (ii), respectively.   
Note that only one neutrino in the first generation contribute to $M_{HOT}$, 
since the electroweak singlet neutrino in the first generation 
has never been in thermal equilibrium in the thermal history of the universe 
(see ref.\cite{solarbbn} for detailed discussion). 
From eq.(9), we can obtain the mass spectrum 
for the values of $\Delta m^2_{{}_{LSND}}=(0.25$--$2.5$) ${\rm{eV}}^2$ 
and $M_{HOT}=(5$--$7$) eV, 
which is shown in Tables I and II.  

The mass spectrum, $m_1\simeq m_3\simeq O({\rm{eV}})$, forces us 
to select the vacuum oscillation solution to the solar neutrino deficit.  
Let us discuss the mass matrix only in the first generation 
at the limit $U_L \rightarrow 1$. 
Because of $S_{{}_{LSND}} \ll 1$, this limit is a good approximation. 
The mass matrix is given by 
%%%%%%%%%%
\barr
\left[ \begin{array}{cc}
\hspace{0.4cm} 0 \hspace{0.3cm}  & \hspace{0.3cm} m_1^{{}_D} \hspace{0.4cm} \\   
\hspace{0.4cm} m_1^{{}_D} \hspace{0.3cm}  & \hspace{0.3cm} M_1 \hspace{0.4cm}
\end{array}\right]
\label{mat1gene}\; . 
\earr
%%%%%%%%
The value of the mixing angle 
is related to the values of the matrix elements in eq.(\ref{mat1gene}). 
The small mixing angle between doublet and singlet neutrinos 
requires the mass matrix to be the `see-saw' type \cite{seesaw}: 
$m_1^{{}_D} \ll M_1$. 
Then, we obtain $m_1 \simeq ({m_1^{{}_D}})^2/M_1 \ll M_1 \simeq m_3
\simeq \sqrt{\Delta m^2_{\odot}}\ll O({\rm{eV}})$.    
Thus, the `small-angle MSW solution' is disfavored. 
On the other hand, the large mixing angle, 
which corresponds to the vacuum oscillation solution, 
requires the mass matrix 
to be almost the `pseudo-Dirac' type \cite{pDirac}: $m_1^{{}_D} \gg M_1$. 
In this case, the value of $m_1$ can be treated as a free parameter, 
if the condition $m_{1} \gg \Delta m_{\odot}^2$ is satisfied, 
and then we can fix $m_1\simeq O({\rm{eV}})$.  
Therefore, the vacuum oscillation solution 
to the solar neutrino deficit 
is selected out.  

The value $m_1 \simeq m_3 \simeq O({\rm{eV}})$ has no conflict 
with the experiments of the neutrino-less double beta decay \cite{dbeta}, 
by which the effective electron neutrino mass is constrained as 
$\langle m_{\nu_e}\rangle < 0.68 \rm{eV}$. 
Since we neglect CP violating phases, i.e. CP is conserved, 
two mass eigenstate, $\nu_1$ and $\nu_3$,  
have opposite CP  eigenvalues $\pm1$, respectively.  
Thus, the cancelation occurs \cite{pDirac}
in the calculation of the effective electron neutrino mass.  
 
We extended the model of Majorana neutrino 
in ref.\cite{dsoscillation} to include the flavor mixing in the first and 
second generations.  
In this simple extended model, the LSND result can be explained 
by the oscillation between the electroweak doublet neutrinos, 
while the solar and atmospheric neutrino deficits can be done by 
the `doublet-singlet oscillation' in the first and  
second generation, respectively, as well as in ref.[5]. 
The requirement for the neutrinos to be the hot dark matter 
selects out only the vacuum oscillation solution 
to the solar neutrino deficit. 

%%%%%%%%%%%%%%%%%%%%%
%  acknowledgments
%%%%%%%%%%%%%%%%%%%%%
The author would like to thank Noriaki Kitazawa for useful comments  
and encouragements.  
This work is supported in part by the Grant in Aid for 
Scientific Research from 
the Ministry of Education, Science and Culture \#H8-1455. 
%
%%%%%%%%%%%%%%%%%%%%%%%%%%%%%%%%%%%%%%%%

%%%%%%%%%%%%%%%%%%%%%%%%%%%%%%%%%%%%%%%%
%
%%%%%%%%%%%%%%%%%%%%%%%%%
%%%%%%%% table %%%%%%%%%%
%%%%%%%%%%%%%%%%%%%%%%%%%
\begin{table}%[h]
\caption{The mass spectrum of type (i)}
\begin{center}
\begin{tabular}{cccc}
\makebox[40mm]{$M_{HOT}\; ({\rm{eV}})$}  & 
\makebox[40mm]{$\Delta m^2_{{}_{LSND}}\;  ({\rm{eV}}^2)$}  & 
\makebox[40mm]{$m_1\simeq m_3\; (\rm{eV})$}  & 
\makebox[40mm]{$m_2\simeq m_4\; (\rm{eV}) $}     \\ \hline 
5   & 0.25   & 1.6 & 1.7  \\
5   & 2.5    & 1.1 & 1.9  \\
7   & 0.25   & 2.3 & 2.4  \\   
7   & 2.5    & 2.0 & 2.5  \\  
\end{tabular}
\end{center}
\end{table}
%%%%%%%%%%%%%%%%%%%%%%%%%
\begin{table}%[h]
\caption{The mass spectrum of type (ii)}
\begin{center}
\begin{tabular}{cccc}
\makebox[40mm]{$M_{HOT}\; ({\rm{eV}})$}  &
\makebox[40mm]{$\Delta m^2_{{}_{LSND}}\;  ({\rm{eV}}^2)$}  &
\makebox[40mm]{$m_1\simeq m_3\; (\rm{eV})$}  &
\makebox[40mm]{$m_2\simeq m_4\; (\rm{eV}) $}     \\ \hline
5   & 0.25   & 1.7 & 1.6  \\
5   & 2.5    & 2.1 & 1.4  \\
7   & 0.25   & 2.4 & 2.3  \\
7   & 2.5    & 2.7 & 2.2  \\
\end{tabular}
\end{center}
\end{table}
%%%%%%%%%%%%%%%%%%%%%%%%%
%%%%%%%%%%%%%%%%%%%%%%%%%
\end{document}